# Analysis of the Phase Transitions in Alkyl-Mica by Density and Pressure Profiles

Hendrik Heinz*[#], Wolfgang Paul[#], Ulrich W. Suter*, and Kurt Binder[#]

[#] Institute of Physics, WA 331, Johannes-Gutenberg-University of Mainz, D-55099 Mainz, Germany

* Institute of Polymers, ETH Zurich, Universitaetsstr. 6, CH-8092 Zurich, Switzerland; correspondence to hendrik.heinz@mat.ethz.ch

# Abstract

In a previous work [H. Heinz, H. J. Castelijns, U. W. Suter, J. Am. Chem. Soc. 115, 9500 (2003)], we developed an accurate force field and simulated the phase transitions in $C_{18}$-mica (octadecyltrimethylammonium-mica) as well as the absence of such transitions in $2C_{18}$-mica (dioctadecyldimethylammonium-mica) between room temperature and 100 °C. Here we analyze (i) average $z$-coordinates of the carbon atoms and interdigitation of the hydrocarbon bilayers, (ii) density profiles and (iii) pressure profiles of the structures along all Cartesian axes. In $C_{18}$-mica, the standard deviation in the $z$ coordinate for the chain atoms is high and more than doubles in the disordered phase. The order-disorder transition is accompanied by a change in the orientation of the ammonium headgroup, as well as decreasing tensile and shear stress in the disordered phase. In $2C_{18}$-mica, the standard deviation in the $z$ coordinate for the chain atoms is low and does not increase remarkably on heating. The backbones display a highly regular structure, which is slightly obscured by rotations in the $C_{18}$ backbones and minor headgroup displacements at 100 °C. Close contacts between the bulky headgroups with sidearms cause significant local pressure which is in part not relieved at 100 °C. An increase of the basal-plane spacing at higher temperature is found in both systems due to larger separation between the two hydrocarbon layers and an increased $z$ spacing between adjacent chain atoms (= decreased tilt of the chains relative to the surface normal), and, in $C_{18}$-mica only, a stronger upward orientation of the $C_{18}$ chain at the ammonium headgroup. The likelihood for chain interdigitation between the two hydrocarbon layers is 24 % to 30 % for $C_{18}$-mica, and 65 % to 26 % for $2C_{18}$-mica (for 20 °C to 100 °C).

 

# I. INTRODUCTION

Organically modified silicates play an important role as additives in polymers to enhance the mechanical and thermal stability, as well as barrier properties in thin films.[1-5] Mica sheets act as mechanically reinforcing additives, which also allow for different colors and optical effects (glittering) of the composite material. Surface modification with hydrocarbon tails is necessary to match the polarity to that of the polymer matrix. Due to the amphiphilic nature of the surfactant bilayer that forms between the modified mica sheets, the organo-micas are also models of a cellular membrane with impenetrable rigid walls.

Previously, we have been able to explain experimental data on phase transitions[6,7] at the interface and within the hydrocarbon chains, and gathered insight into the chain conformations by carrying out simulations.[8,9] We present now an analysis of the structural transitions, with emphasis on coordinate fluctuations in the chains and pressure profiles, using the data from our simulation.[8] Representative snapshots of the systems are depicted in Fig. 1. We consider mica with two common organic cations, which we designate as "$C_{18}$" (octadecyltrimethylammonium ion, $n\text{-}C_{18}H_{37}N(CH_3)_3^+$), and "$2C_{18}$" (dioctadecyldimethylammonium ion, $(n\text{-}C_{18}H_{37})_2N(CH_3)_2^+$). The $C_{18}$-mica system undergoes two order-disorder transitions on heating, one involving intra-chain disorder, and the other involving disorder of the ammonium headgroups on the mica surface. The $2C_{18}$-mica system exhibits no significant changes due to geometrical restraints; the number of *gauche*-incidences in the chain backbones remains at an overall low fraction at higher temperature.[8] In the present paper, we consider 100 % exchange of alkali ions; the phase transitions on heating for variable degrees of ion exchange and other chain lengths proceed analogously.[9]

In section II, we summarize the MD simulation method used. In section III, we evaluate the average *z* coordinates of the carbon atoms, analyze the changes of the basal-plane spacings with temperature, and discuss chain interdigitation. In section IV, the density profiles and pressure profiles of our system along the Cartesian axes are presented. We point out their relation to structural changes and phase transitions. Section V comprises conclusions.

## II. DETAILS OF THE SIMULATION

We carried out molecular dynamics, using a chemically realistic, atomistic model. The simulation boxes are periodic and consist of 4×3×1 unit cells of muscovite mica $2M_1$ and 24 intercalated alkylammonium ions,[8] equivalent to the dimensions 2.0767 nm × 2.7046 nm × $z$. To allow for equilibration of the basal-plane spacing, the periodicity in $z$-direction is eliminated by choosing the box length approximately 5 nm beyond the extension of the system. The force field is based on the consistent force field 91 (cff91),[10] which was augmented for the simulation of aluminosilicates.[8, 11] The potential energy consists of terms for bond stretching (quadratic, cubic, and quartic), angle bending (quadratic, cubic, and quartic), torsions (three-term), cross terms, electrostatic interactions, and a 9-6 Lenard-Jones potential.[10] We surmise that the cross terms are negligible, also cubic and quartic contributions to bonds and angles could be avoided. Moreover, modeling of the dispersive interactions with an $r^{-4}$ distance dependence might improve the performance.[12] In spite of these possible improvements, our augmented cff91 fares well and both the hydrocarbon part and the mineral part are reproduced within few percent of geometric deviations. We simulated the NVT ensemble, using the cell multipole method for electrostatics, Verlet's integration scheme, and velocity scaling for temperature control. The time step was chosen as 1 fs and we simulated the system for a duration >500 ps. This length of time proved to be sufficient for equilibration, and 100 snapshots at an interval of 1 ps are taken to carry out the analysis of the system.

## III. CARBON BACKBONES AND INTERDIGITATION

We have computed the $z$ coordinates of the C and N atoms for an average pair of alkylammonium ions forming a bilayer (Figures 2 and 3). An average over 100 snapshots is taken, each containing 12 pairs of chains, so that each data point is the average over 1200 values. Standard deviations for the individual atoms are included. The standard deviations of the average coordinates, however, are only $1/\sqrt{1200} = 1/35$ of the values indicated by the error bars.

III.A. $C_{18}$-mica

The sequence of the C and N atoms in the order of increasing $z$ coordinate is shown in Fig. 2. Two methyl carbons are followed by the nitrogen; then the carbon atoms of the $C_{18}$ chain ensue. The third methyl group is located after the first chain atom or directly after the nitrogen (see the arrows in Fig. 2). We find an increase of the basal-plane spacing with increasing temperature, which is composed as follows:

(a) The average gap between the two hydrocarbon layers is increasing from $150 \pm 5$ pm to $250 \pm 10$ pm.

(b) The average intra-chain difference in $z$ coordinate between two neighboring carbons in the backbone increases slightly from 61.6 pm towards 62.2 pm. Considering the 17 C–C bonds in each of the two alkyl backbones, this leads to an overall increase in basal-plane spacing of ~20 pm.

(c) An important change occurs at the two N–C bonds connecting the ammonium headgroups with the $C_{18}$ chains: The total contribution to the $z$ coordinate increases from $60 \pm 3$ pm to $105 \pm 3$ pm while the N−C bonds to the $C_{18}$ backbone change their average tilt angle to the surface normal from 79° towards 70° (see Fig. 3).

With these three main factors for the increase in basal-plane spacing, the total difference amounts to $165 \pm 12$ pm from 20 °C up to 100 °C. It is in agreement with the experimental result of $130 \pm 20$ pm that refers to the increase in basal-plane spacing from 20 °C up to 70 °C.[6] The original graph in ref. 6 suggests a further increase with temperature, in accordance with our computed value.

A conspicuous feature are the standard deviations of the individual $z$ coordinates (see Fig. 2). They measure up to ±150 pm at 20 °C at the ends of the chain, lowering to ±60 pm near the ammonium head group. At 100 °C, the fluctuation is significantly higher towards the end of the chain with standard deviations of the $z$ coordinate of ±340 pm, lowering to ±70 pm near the ammonium head group. The nitrogen atom is rather tied in both cases, with only ±15 pm or ±23 pm fluctuations, respectively. The likely reason is its role as a center of strong electrostatic-repulsive interaction with the mica lattice. The substantial change in vertical flexibility of the backbone atoms increases both the probability for interdigitation as well as the probability for large separations between the chain ends at higher temperature.

We define interdigitation of the two hydrocarbon layers in their $z$ coordinates as overlap of the C-terminal carbon atoms in their $z$ coordinates. Interdigitation is not found as an average over all chain backbones at both temperatures (Fig. 2). For accurate estimates, we define the probability of interdigitation, assuming Gaussian distributions $P_1(z)$ and $P_2(z)$ of the $z$ coordinates of both chain ends with the same standard deviation $\sigma_z$ for the two ends (the standard deviation appears not exactly the same way in Fig. 2 because the $z$ coordinate is calculated relative to the bottom basal-plane so that fluctuations in the upper layer are damped by those in the lower layer). The two mean values are separated by the gap $\Delta z$:

$$P_2(z) = P_1(z - \Delta z). \tag{1}$$

In this case, $P_2$ has the higher mean, and the probability of interdigitation $p_i$ is given by

$$p_i = \int_{-\infty}^{+\infty} P_2(z_2) \left( \int_{z_2}^{+\infty} P_1(z_1) dz_1 \right) dz_2. \tag{2}$$

The probability for interdigitation by a minimum value of $\delta$ is then given as

$$p_i(\delta) = \int_{-\infty}^{+\infty} P_2(z_2) \left( \int_{z_2+\delta}^{+\infty} P_1(z_1) dz_1 \right) dz_2. \tag{3}$$

Using $\sigma_z = \pm 150$ pm and $\Delta z = 150$ pm at 20 °C, we obtain $p_i(0) = 0.24$ and $p_i(150 \text{ pm}) = 0.08$. Inserting $\sigma_z = \pm 340$ pm and $\Delta z = 250$ pm at 100 °C, we obtain $p_i(0) = 0.30$ and $p_i(150 \text{ pm}) = 0.20$. These values demonstrate that, at 20 °C, the probability for any interdigitation (0.24) is far higher than the probability for more than 150 pm interdigitation (0.08). At 100 °C, both probabilities are increased, and the probability for any interdigitation (0.30) remains only moderately higher than the probability for more than 150 pm interdigitation (0.20). These remarkable changes in backbone flexibility as well as the modified headgroup orientation are in concordance with phase transitions involving both intra-chain and headgroup disordering.

III.B 2$C_{18}$-mica

The sequence of C atoms is shown in Fig. 4. At the highest and the lowest $z$ coordinate, the two methyl groups of the ammonium headgroups appear, followed by the nitrogen atom. The next three atoms at almost the same height are the N-terminal carbon atoms of the bent $C_{18}$ chains (Fig. 1). They are ensued by 15 pairs of atoms at similar altitude, which belong to both $C_{18}$ backbones. Finally, we find the three remaining single C atoms of each outstretched single $C_{18}$ strand. The last atoms of each hydrocarbon layer are interdigitated on average at 20 °C, while they move apart from each other on increase of the temperature (Fig. 4). The increase in basal-plane spacing is composed of two major factors:

(a) At 20 °C, we find an average net interdigitation of $28 \pm 2$ pm, whereas at 100 °C the layers are separated on average by $64 \pm 3$ pm. These closest approaches occur between the outstretched $C_{18}$ arms (see Fig. 1).

(b) The C–C distances in $z$ direction within the backbones display an interesting two-step pattern (see Fig. 4). At 20 °C, they vary in the order 140 pm – 80 pm – 140 pm – 80 pm – ⋯, whereas at 100 °C the alternation is weaker: 120 pm – 110 pm – 120 pm – 110 pm – ⋯. The effective gain in $z$ coordinate on increasing temperature is thus 5 pm in each of the 17 C–C bonds of each layer and contributes $170 \pm 20$ pm to the gain in basal-plane spacing at 100 °C.

Accordingly, the increase in layer separation of ~90 pm and the erection of the chain backbones of ~170 pm yield a total increase in basal-plane spacing of $260 \pm 20$ pm (no literature value reported for 100 % ion exchange). In contrast to $C_{18}$-mica, we find no structural changes near the head groups.

Strikingly, the standard deviations of the $z$ coordinates of the individual atoms are several times smaller than in $C_{18}$-mica at both temperatures and increase less than 40 % on heating (see Fig. 4). At the C-terminus of the chain, the atoms fluctuate ±51 pm at 20 °C and ±69 pm at 100 °C, near the head group these fluctuations are only ±20 pm and ±25 pm, respectively. As in $C_{18}$-mica, the nitrogen atoms are tied most strongly, although the nearby carbon atoms in $2C_{18}$-mica also have little flexibility (Fig. 4). Since the overall degree of disorder is low, a substantial order-disorder transition in the chain backbones can be ruled out.

The exact probabilities for interdigitation according to Eq. (3) are $p_i(0) = 0.65$ at 20 °C and $p_i(0) = 0.26$ at 100 °C. Deep interdigitation is not possible due to the geometric restraints of the system (see Fig. 1).

## IV. DENSITY AND PRESSURE PROFILES

Density profiles and local pressures have proved to be useful indicators of phase transitions in lipid monolayers[13, 14] and single crystals,[15] in the formation of lipopolymer superstructures,[16] to analyze confined fluids[17] and local dynamics in glasses.[18] We present the density and pressure profiles of our systems in Figures 5 and 6. The pressures are calculated as an average over the entire cross-sections of the box $A_\alpha$ along $x$, $y$, and $z$ coordinate according to[19]

$$p_{\beta\alpha}(\alpha') = \frac{1}{2A_\alpha \Delta\alpha} \left\langle \sum_{\alpha'-\Delta\alpha \leq \alpha_i \leq \alpha'+\Delta\alpha} m_i v_{i\beta} v_{i\alpha} \right\rangle + \frac{1}{2A_\alpha} \left\langle \sum_{i=1}^{N} F_{i\beta}\, \text{sgn}(\alpha_i - \alpha') \right\rangle. \qquad (4)$$

The interval is chosen as $2\Delta\alpha = 3$ pm, so that on average two to five atoms are located within the volume elements $2A_\alpha \Delta\alpha$ in each of the 100 snapshots. Thus, we obtain between 700 and 2000 data points for each graph, depending on the extension of our system in the respective direction, with each data point representing an average over several hundred atoms.

In general, the kinetic contribution to the pressure tensor (the first term in Eq. (4)) is on average 20 to 40 MPa for the diagonal elements, only the values of $p_{zz}^{kin}(z)$ in the open $z$ direction fluctuate strongly and are counterbalanced by the interatomic component $p_{zz}^{int}(z)$ (the second term in Eq. (4)). All off diagonal elements of the kinetic part of the stress tensor are small and fluctuate structurelessly around zero. The dominant part in our condensed matter systems is the interatomic part of the pressure tensor (the second term in Eq. (4)). We note in general that an average pressure component of ~200 MPa acting on a plane perpendicular to the $x$ or $y$ axis would cause no more than 1.5 % distortions in geometry because the elastic moduli of mica are ~70 GPa, and >20 % of the cross-section along $x$ or $y$ consists of mica.

IV.A $C_{18}$-mica

Due to significant flexibility of the structures already at room temperature, the density and pressure profiles (especially along the $z$ direction) are very complex. The density profiles along the $x$ and $y$ coordinate are dominated by the constituting atoms of mica at well-defined positions and the peaks broaden with increasing temperature due to larger oscillations (Fig. 5). Along the $z$ coordinate, the curve is essentially structureless in the hydrocarbon part. Even at 20 °C the coordinate fluctuations of the individual atoms are high (cf. Fig. 2), and the roughly constant value of $d(z)$ in the hydrocarbon moiety is in agreement with a largely disordered phase. The diagonal elements $p_{xx}$ and $p_{yy}$ of the pressure tensor have a mean value of $-150$ MPa at 20 °C. This tendency towards a light contraction is lost at 100 °C, the value becomes roughly $+40$ MPa. The off-diagonal elements and $p_{zz}$ are either near zero at room temperature or approach this value at 100 °C within the accuracy of $\pm 10$ MPa (Fig. 5). For example, $p_{zx}$ still had the noticeable value of $p_{zx} \approx 110$ MPa at 20 °C, while at 100 °C the value is almost zero. Most of the stress is relieved on increase of the temperature, with the help of headgroup rearrangements. This observation is in concordance with two melting transitions, one involving disordering of the backbones, and the other rearrangements of the headgroups as well.[8]

IV.B $2C_{18}$-mica

Also here, the density profiles along $x$ and $y$ mainly indicate the mica structure because the hydrocarbon chains have more flexibility than the mica lattice and do not yield intense peaks. Along the $z$ axis, we recognize the structural two-step pattern at 20 °C as pointed out in section III.B. At 100 °C, $d(z)$ still indicates a fairly ordered structure, although the curve is more uniform and the intervals between the C atoms are alternating less. No significant order-disorder transition can be assigned. $p_{xx}$ is $-115$ MPa at 20 °C and relieved partly to $+50$ MPa at 100 °C. $p_{yy}$ decreases from +190 MPa to zero, while $p_{zz}$ assumes a value around 80 MPa at both temperatures (see Fig. 6). We note here the influence of the electrostatic forces from the planar mica surfaces, which enforce the headgroup arrangement of the 2C$_{18}$ ions. Mica acts like a template and leads to sterically demanding structures with varying degree of local stress. The bulky ammonium headgroups with the C$_{18}$ sidearm are densely packed along the *y* coordinate at room temperature and loosely alongside *x* (Fig. 1), which leads to an eclipsed arrangement of the organic chains. The associated stress can be partly relieved by minor headgroup displacements and rotations in the chain backbones at higher temperature. These changes help to shift $p_{xx}$, $p_{yx}$, $p_{xy}$, and $p_{yy}$ towards zero (Fig. 6). While stress in the *xy* plane can be counterbalanced at higher temperature by chain rotations and small headgroup displacements, shear forces in *z* direction ($p_{zx}$ and $p_{zy}$) are even increased up to +170 MPa (Fig. 6) because the alkyl chains cannot shear along the *z* direction due to the rigidity of the mica sheets.

# V. CONCLUSIONS

The standard deviations of the carbon $z$ coordinates are a valuable order parameter for organo-micas. We found that these fluctuations in $C_{18}$-mica are up to five times as much as in $2C_{18}$-mica. The significant increase of the coordinate fluctuation on heating for $C_{18}$-mica ($\pm 150$ to $\pm 340$ pm) is in agreement with a phase transition involving the conversion of a slightly disordered backbone into a conformationally disordered structure. Also, the ammonium headgroups change their average orientation so that the first carbon atom of the $C_{18}$ backbone points more upward at higher temperature. From previous experimental measurements[6,7] and the simulation,[8] we found that two transitions occur in this temperature range and the second transition involves headgroup rearrangements. Concomitantly, we find a uniform density profile $d(z)$ and relief of tensile and shear stress in the disordered phase. The probability of interdigitation between the two hydrocarbon layers amounts to 24 % at room temperature and increases to 30 % at 100 °C.

In $2C_{18}$-mica, the standard deviations of the $z$ coordinates are relatively small ($< \pm 70$ pm) and a regular ordering of the chain backbones is apparent at both temperatures. The headgroups are voluminous due to the second $C_{18}$ sidearm, and cause local stress at room temperature. At elevated temperature, the tensile and shear stress components in the $xy$ plane can be largely reduced by minor headgroup displacements (not across cavities) and rotations in the chain backbone. Thereby, the highly regular two-step pattern along the chain backbones is diluted. The shear forces along the $z$ coordinate increase with higher temperature. The probability of interdigitation between the hydrocarbon layers amounts to 65 % at room temperature and 26 % at 100 °C.

The basal-planes spacing increases at higher temperature in both $C_{18}$-mica and $2C_{18}$-mica. This increase, thus, is not necessarily connected with a phase transition. The main contributions are an increased gap between the two hydrocarbon layers, and a more or less pronounced increase of the $z$ separation between the backbone atoms (equals a decreased tilt angle of the chains relative to the surface normal). In the case of $C_{18}$-mica, a significant contribution arises from an upward orientation of the $C_{18}$-backbones directly at the ammonium group at higher temperature.

**Acknowledgment**

We are grateful for helpful discussions with Prof. Sanjay Puri, Jawaharlal Nehru University, Delhi. The support from the Swiss National Science Foundation, ETH Zurich, University of Mainz, and the Studienstiftung des Deutschen Volkes is gratefully acknowledged.

**Figure Captions**

FIG. 1. Molecular dynamics snapshots of equilibrated mica-alkyl-alkyl-mica structures, viewed along the $x$ axis (ref. 8). (a) $C_{18}$-mica, 20 °C. (b) $C_{18}$-mica, 100 °C. (c) $2C_{18}$-mica, 20 °C. (d) $2C_{18}$-mica, 100 °C. Hydrogen atoms are omitted in the chain backbones. C atoms are shown as sticks, N atoms of the headgroups as small green balls.

FIG. 3. Average $z$ coordinates of C and N for a pair of $C_{18}$ ions on mica forming a bilayer. The two residues are distinguished by filled and empty circles, respectively. Two methyl groups of the $(CH_3)_3N(C_{18}H_{37})^+$ ions are located near the mica sheets, followed by the N atom. The third methyl group moves closer to the nearest mica sheet at higher temperature (emphasis by arrows), lifting the first chain atom.

FIG. 3. Approximate orientation of the ammonium headgroup in $C_{18}$-mica at 20 °C (left) and at 100 °C (right). The hydrocarbon chain shows a tendency to move up relative to the mica surface, while the nitrogen atom remains at the same height.

FIG. 4. Average $z$ coordinates of C and N for a pair of $2C_{18}$ ions on mica forming a bilayer. The two residues are distinguished by filled and empty circles, respectively. The interface region of the two layers is zoomed in to show the slight interdigitation at room temperature.

FIG. 5. Pressure profiles (Pa) and density profiles (kg/m$^3$) through the entire box of $C_{18}$-mica along all coordinates (nm). The graphs are shown at 20 °C (black) and 100 °C (grey).

FIG. 6. Pressure profiles (Pa) and density profiles (kg/m$^3$) through the entire box of $2C_{18}$-mica along all coordinates (nm). The graphs are shown at 20 °C (black) and 100 °C (grey).

FIG. 1

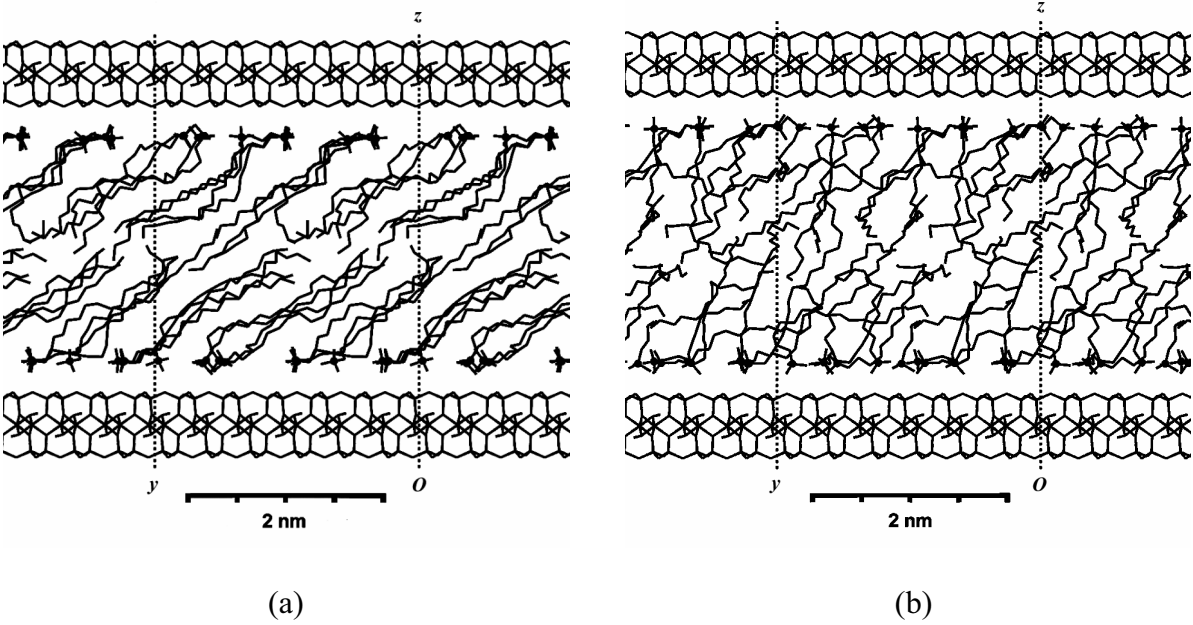

(a) (b)

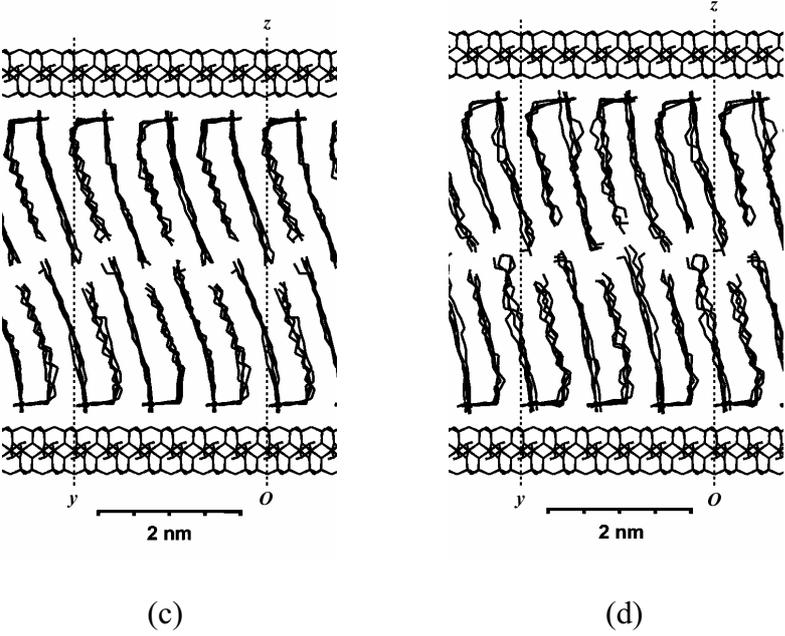

(c) (d)

FIG. 2

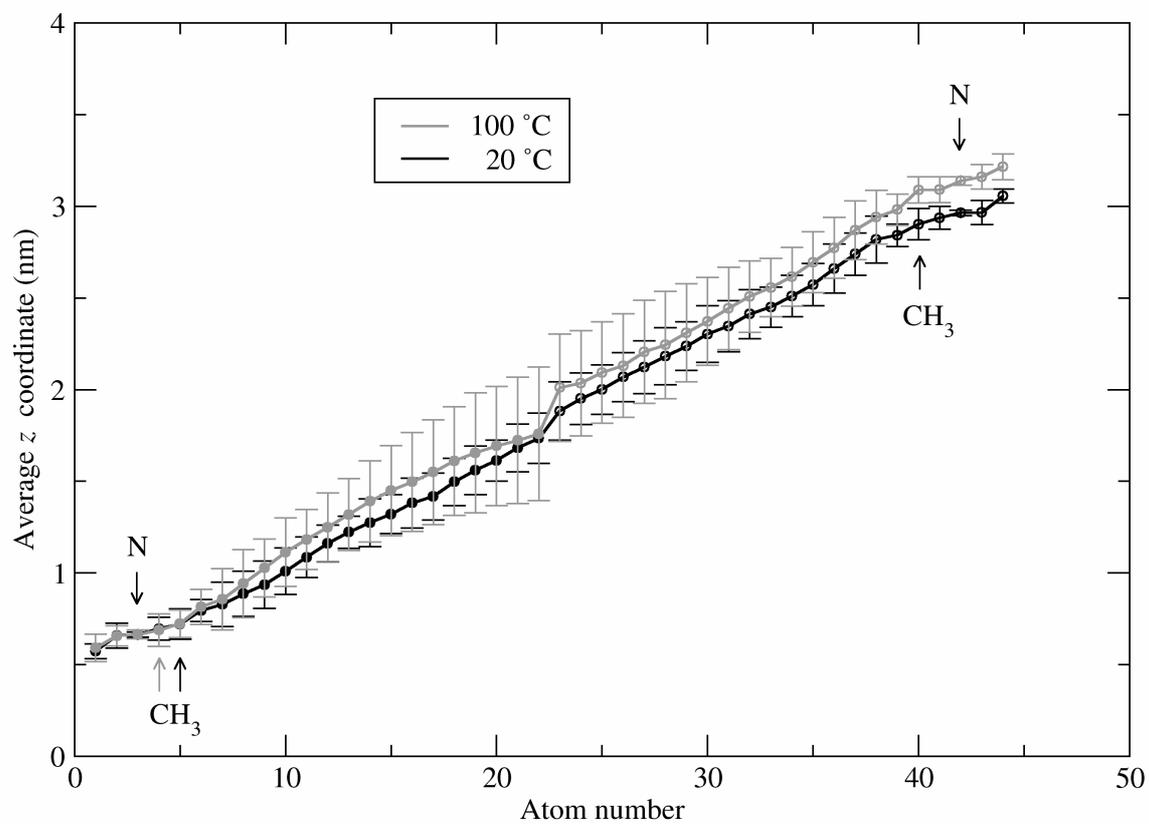

FIG. 3

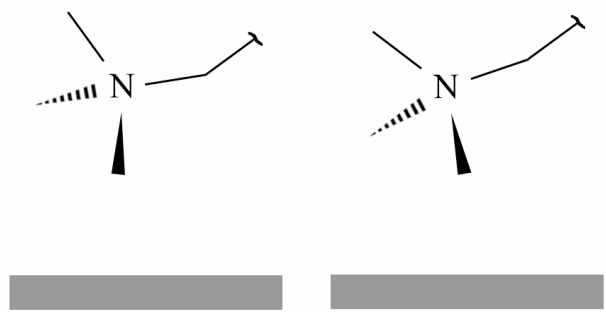

FIG. 4

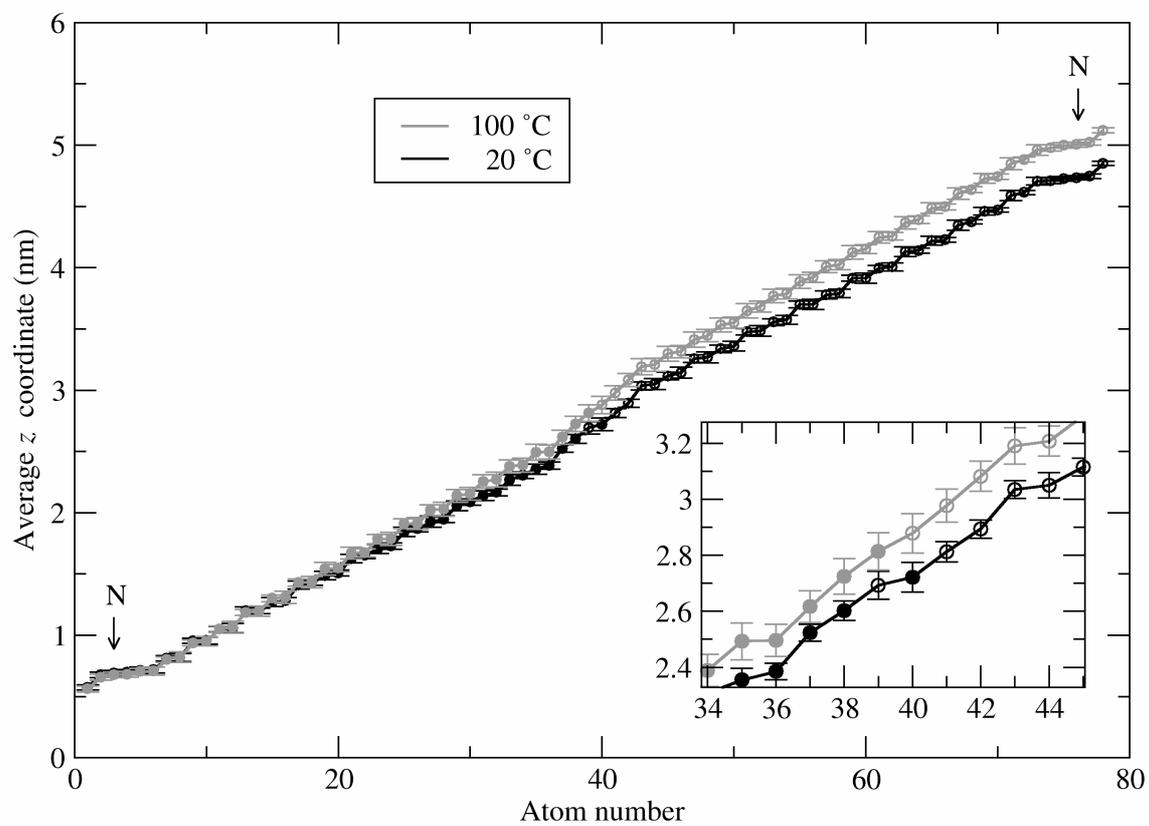

                                            

FIG. 5

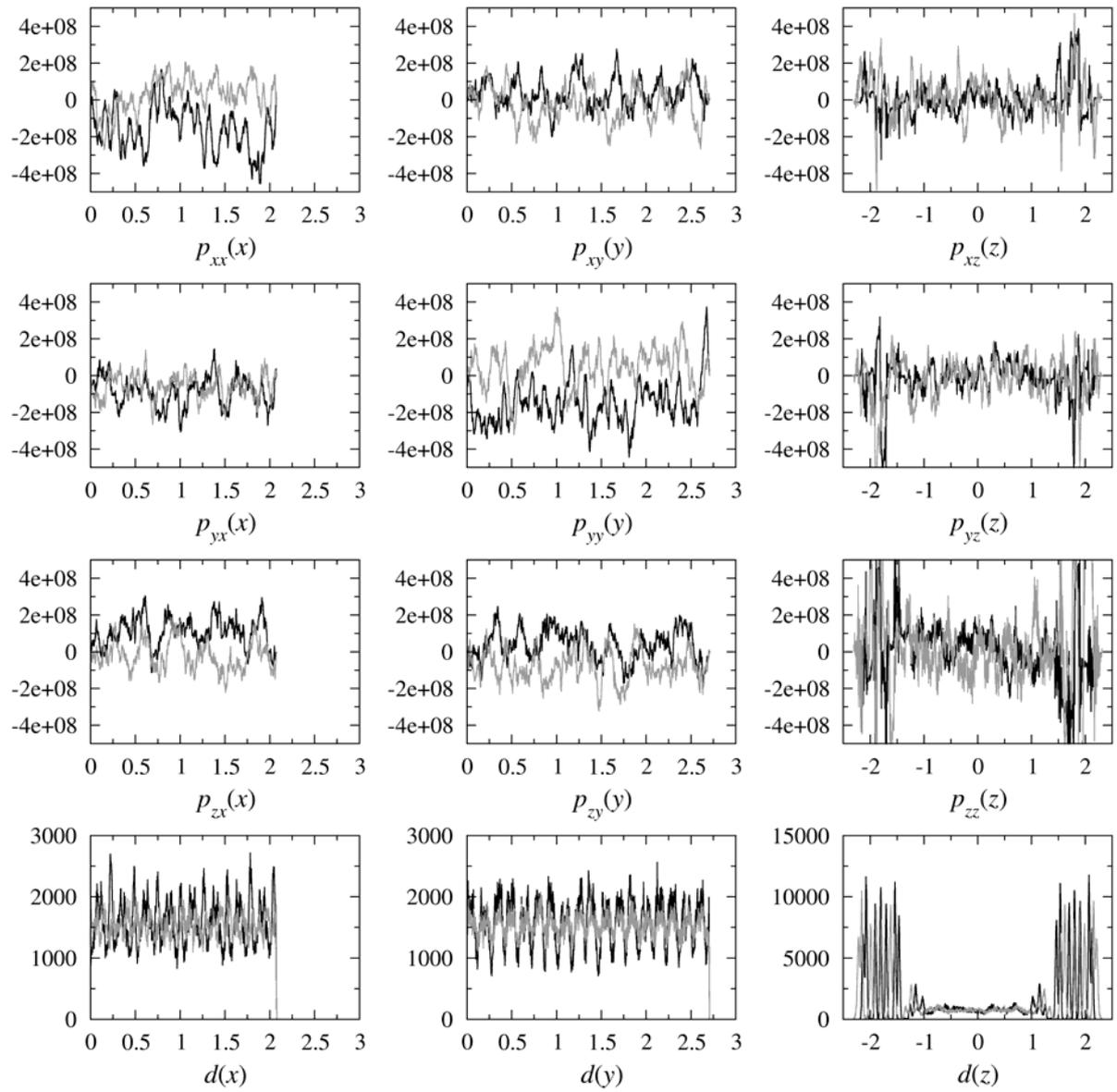

 mica_analysis_revised5

FIG. 6

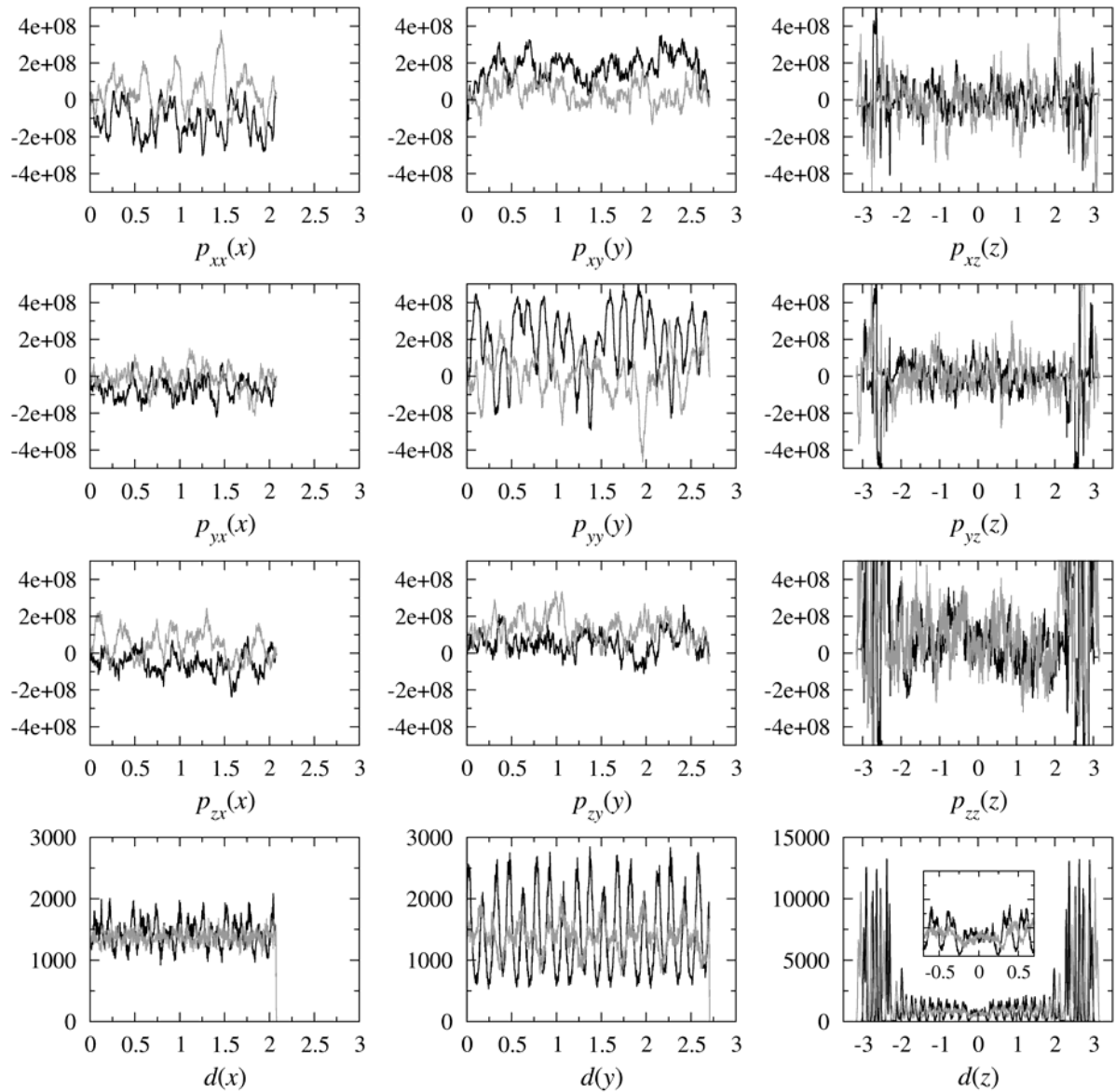